\documentstyle[twocolumn,aps,prl]{revtex}

\input epsf.sty

\begin{document}

\wideabs{
\draft
\title{Strange particle production at RHIC in a single-freeze-out model
\cite{grant}}
\author{Wojciech Broniowski and Wojciech Florkowski}
\address{The H. Niewodnicza\'nski Institute of Nuclear Physics,
PL-31342 Cracow, Poland}
\maketitle
\begin{abstract}
Strange particle ratios and $p_\perp$-spectra are calculated in a thermal model
with single freeze-out, previously used successfully to describe non-strange
particle production at RHIC. The model and the recently released
data for $\phi$, $\Lambda$, $\overline{\Lambda}$ and $K^*(892)^0$ are
in very satisfactory agreement, showing that the thermal approach can be used
to describe the strangeness production at RHIC.
\end{abstract}
}

\section{Introduction}

In this paper we calculate the ratios and the $p_\perp$-spectra for
the strange particles produced at RHIC at $\sqrt{s}=130$ GeV~A.  We
use the thermal model with {\em single} freeze-out. The work is a
direct follow-up of our previous study reported in Ref. \cite{wbwf}.
The very recent data from the STAR collaboration on the production of
$\phi$ \cite{yama}, $\Lambda$ \cite{starLambda}, $\overline{\Lambda}$
\cite{starLambda}, and $K^*(892)^0$ \cite{starKstar} are confronted
with the model and a very good agreement is found.

Enhanced strangeness production in the high-energy nuclear collisions,
relative to more elementary $p p$ or $e^+ e^-$ collisions, was
proposed many years ago \cite{rafmul,kochmulraf} as a signal of the
quark-gluon plasma formation (for more recent arguments and discussion
see, {\it e.g.}, \cite{raf}). In the meantime, the experimental
evidence for the enhancement has indeed been found, however its
significance for the plasma formation remains an open problem
\cite{singh}.  The ratios of the particle abundances measured in Pb+Pb
collisions at SPS ($\sqrt{s}=$17 GeV A) and in Au+Au collisions at
RHIC ($\sqrt{s}=$130 GeV A) are very well explained in the framework
of the thermal models which use the hadronic degrees of freedom only
\cite{pbmsps,pbmrhic,wfwbmm}. In this approach one assumes that the
net strangeness is zero, and the strange hadrons are in a complete
thermal and chemical equilibrium with other hadrons. A combination of
the thermal model with a suitable hydrodynamic expansion (the {\it
single-freeze-out model} of Ref. \cite{wbwf}) led to a very good
quantitative description of the RHIC transverse-momentum spectra of
pions, kaons, and protons.  In this paper we present the predictions
of this model for the production of other particles, with a special
emphasis on the strangeness production. Without any refitting of the
two thermal and two expansion parameters of Ref.~\cite{wbwf}, we
calculate the $p_\perp$-spectra of $\phi$, $\Lambda$,
$\overline{\Lambda}$, $K^*(892)^0$, $\Xi$, $\Sigma$, and $\Omega$. The
spectra of $\phi$, $\Lambda$, $\overline{\Lambda}$ and $K^*(892)^0$
are in a very good agreement with the very recently released data
\cite{yama,starLambda,starKstar}. The spectra of other particles are
also presented and will be confronted with the incoming future
data. We also compute the thermal spectrum of the $J/\psi$.

\section{Definition of the model}

The main assumption of our model is that the chemical freeze-out
occurs simultaneously with the thermal freeze-out, i.e., the hadrons
decouple completely when the thermodynamic parameters reach the
freeze-out conditions. In other words, the possible elastic
rescattering processes after the chemical freeze-out are neglected.
This assumption opposes the most popular scenario \cite{heinzr}, where
the two freeze-outs are separated. The argument used in this context
is the difference of scales, following from the fact that the elastic
cross sections are, for most channels, much larger than the inelastic
cross sections (note, however, that this is not the case for $p
{\overline p}$ interactions, where the inelastic channel
dominates). The scenario with a single freeze-out is natural if the
hadronization occurs in such conditions that neither elastic nor
inelastic processes are effective. An example here is the picture of the
supercooled plasma of Ref. \cite{rafelski}.  

Recently, new hints have been presented in favor of smaller
rescattering effects at RHIC: a successful reconstruction of the
$K^*(892)^0$ states has been achieved by the STAR Collaboration
\cite{starKstar}. As indicated by the authors, either the daughter
particles from the decay $K^*(892)^0 \rightarrow \pi K$ do not
rescatter or the expansion time between the chemical and thermal
freeze-out is shorter than the $K^*(892)^0$ lifetime ($\tau$ = 4
fm/$c$). In addition, the fact that the experimentally measured yields
of $K^*(892)^0$ \cite{starKstar} are very well reproduced in the
framework of the thermal models \cite{pbmrhic,wfwbmm} suggests the
picture with a very short expansion time between the two freeze-outs,
as proposed in Ref. \cite{wbwf}.


An important feature of our analysis is a {\em complete} treatment of
the hadronic resonances, with all particles from the Particle Data
Table \cite{PDG} taken into account in the analysis of both the ratios
and the spectra.  This effectively leads to ``cooling'' of the
hadronic spectra by 30-40MeV \cite{wfwbmm}, which is a very important
effect \cite{hagedorn}, crucial to obtain agreement with the data.

The final ingredient of our model is the choice of the freeze-out
hypersurface, which is, in the spirit of
Ref.~\cite{bjorken,baym,Kolya,siemens,SSH,BL,Rischke,SH}, defined by
the condition
\begin{equation}
\tau = \sqrt{t^2-r^2_x-r^2_y-r^2_z} = \hbox{const}.
\label{tau}
\end{equation}
The transverse size, $\rho=\sqrt{r_x^2+r_y^2}$, is limited by the
condition $\rho < \rho_{\rm max}$. The $t$ and $r_z$ coordinates,
appearing in the boost-invariant combination, are not limited, hence
the model is boost-invariant. We have checked numerically that this
approximation works very well for calculations in the central-rapidity
region.  Finally, we assume that the four-velocity of the hydrodynamic
expansion at freeze-out is proportional to the coordinate (Hubble-like
expansion),
\begin{equation}
u^{\mu } =\frac{x^{\mu }}{\tau }=\frac{t}{\tau }\left(
1,\frac{r_{x}}{t},\frac{r_{y}}{t},\frac{r_{z}}{t}\right).
\label{umu}
\end{equation}

The question arises to what extent the assumptions
(\ref{tau},\ref{umu}) are realistic. Typically, in a hydrodynamic
approach the freeze-out hypersurface contains, in the $\rho$-time
plane, a time-like part, and a space-like part
\cite{baym,Kolya,siemens,SSH,BL,Rischke,SH}.  There is a conceptual
problem here \cite{bugaev,csernai,neymann,magas}.  A part of the
particles emitted from the space-like surface goes backwards into the
firecylinder and re-equilibrates. A common prescription is to exclude
these contributions by hand.  In our parametrization we neglect the
space-like part altogether, thus avoiding the above problem.  The
time-like part of the hypersurface has, in many calculations, the
feature that the outer regions in the transverse direction freeze
earlier than the inner regions. This is opposite to what follows from
Eq. (\ref{tau}), or from the commonly used versions of the blast-wave
model, where the freeze-out occurs at a constant value of time in the
$\rho$-time plane.  Our present form for the hypersurface and for the
velocity field corresponds to the so called {\em scaling solution}
\cite{baym,CF2,biro}, which is obtained in the case of a small sound
velocity in the medium.  Naturally, validity of the assumptions and
their relevance for the results should be examined in a greater
detail. In particular, the consequences of the specific choice of the
freeze-out hypersurface must be studied. The spherically symmetric
case without the decays of resonances was investigated in
Ref. \cite{lhs}.  In Ref. \cite{wbwf} we have checked that two
different models lead to very close predictions for the spectra. Other
parametrizations may be also verified with the help of the formulas
given in the Appendix.  One should bare in mind that at the moment
there is no microscopic approach capable of reproducing all the
features of the RHIC data (abundances, spectra, and HBT radii), such
that the issue of how the hadronization and freeze-out occur is very
much open.

The model has four parameters possessing clear physical
interpretation. The first two parameters are the temperature, $T$, and
the baryon chemical potential, $\mu_B$. These are fixed by the
analysis of the ratios of the particle abundances \cite{wfwbmm}.  The
next two parameters, concerning the model of expansion, are the
invariant time, $\tau$, and the transverse size, $\rho_{\rm max}$.
The invariant time describes the lifetime of the system, and $\tau^3$
controls the overall normalization of the hadron multiplicities.  On
the other hand, the ratio $\rho_{\rm max}/\tau$ influences the slopes
of the $p_\perp$-spectra.  The expansion parameters have been fitted
to the spectra \cite{velko,harris} in Ref. \cite{wbwf} by the
least-square method.  We note that $\rho_{\max}$ is directly related,
through Eq. (\ref{umu}), to the amount of the transverse flow, crucial
for the shape of the transverse-momentum distributions. 
The maximum value of the transverse fluid velocity, reached at the boundary, 
is $\beta_\perp^{\rm max}=0.66$, while the average value is 
$\langle \beta_\perp \rangle = 0.49$.

We use the previously-found
values \cite{wfwbmm} for the thermal parameters:
\begin{equation}
T = 165 \hbox{ MeV}, \hspace{2cm} \mu_B = 41 \hbox{ MeV}.
\label{tandmu}
\end{equation}
The fit to the $p_\perp$ spectra of pions, kaons and
protons for the most central collisions yielded \cite{wbwf}
\begin{equation}
\tau = 7.66 \hbox{ fm}, \hspace{2cm} \rho_{\max} = 6.69  \hbox{ fm}.
\label{tauandrho}
\end{equation}

In the present paper we use the values of the parameters given by
Eqs. (\ref{tandmu}) and (\ref{tauandrho}), and calculate the spectra
of other hadrons. As in  Ref. \cite{wbwf}, the spectra are
obtained from the Cooper-Frye  \cite{CF2,CF1} formula
\begin{equation}
\frac{dN_{i}}{d^{2}p_{\perp }dy} =
\int p^{\mu }d\Sigma _{\mu }\ f_{i}\left(p\cdot u\right) ,
\label{Ni}
\end{equation}
where $p^{\mu }$ is the four-momentum of the particle, $d\Sigma _{\mu
}$ is the volume element of the hypersurface defined by condition
(\ref{tau}), and $f_{i}$ is the phase-space distribution function for
particle species $i$. It is composed from the initial and secondary
particles, see the Appendix, proving that this is a correct approach
for the case where $d\Sigma_\mu \sim u_\mu$. The formulas valid for a
general expansion are also given in the Appendix.

\section{Results}

The thermodynamic parameters used in our model, Eq. (\ref{tandmu}),
yield the following ratios of the strange hadron abundances:
\begin{equation}
{\Omega^- \over \Xi^-} = 0.18, \,\,
{\Xi^- \over \Sigma^-} = 0.55, \,\,
{\Sigma^- \over \Lambda} = 0.20, 
\label{ratios1}
\end{equation}
\begin{equation}
{\Lambda \over p} = 0.47 \hspace{0.4cm} 
(0.49 \pm 0.03)_{\rm expt.},
\label{ratios12}
\end{equation}
\begin{equation}
{\Omega^+ \over \Omega^-} = 0.85, \,\,\,
{\Xi^+ \over \Xi^-} = 0.76, \,\,\,
{\Sigma^+ \over \Sigma^-} = 1.02.
\label{ratios2}
\end{equation}
The experimental value for the $\Lambda/p$ ratio follows from the data published in 
Refs. \cite{starLambda,starantip}.
The feeding from weak decays has been included in the above
ratios. For example, the multiplicity of $\Xi$ contains a contribution
from the $\Omega$ decay, and the multiplicity of $\Lambda$ contains
contributions from the decays of $\Xi, \Sigma$, and (indirectly)
$\Omega$.  When the feeding from weak decays is excluded, the
following ratios are noticeably altered:

\begin{equation}
{\Sigma^- \over \Lambda} = 0.36, \,\,\,\,\,\,
{\Lambda \over p} = 0.26.
\label{ratios1nw}
\end{equation}
The abundance of the $\Xi$ hyperon is sensitive to the inclusion of
higher $\Xi$ resonances, in particular $\Xi(1690)$, whose properties
(spin) are not known. This leads to a theoretical error of at least
10\%. This is a general feature of the thermal fits. The heavier
particles receive contributions from their excited states, however the
experimental particle spectrum is less known as the mass increases.
The higher mass states are suppressed by the thermodynamic factor, on
the other hand, they are more numerous, according to the Hagedorn
hypothesis.  For the recent discussion of this issue see
Refs. \cite{hag1,hag2}.

For the $J/\psi$ we obtain
\begin{equation}
{J/\psi \over \pi^-} = 0.12 \times  10^{-6}.
\label{ratios3}
\end{equation}
The abundance of $J/\psi$ is an order of magnitude lower than the
values obtained in Refs. \cite{gaz1,gaz2}, where a wide class of
reactions is studied and the ratio $(J/\psi)/h^- \simeq 10^{-6}$ is
found. Note, however, that due to the large mass of $J/\psi$, the
ratio is very sensitive to the chosen freeze-out temperature.  The
ratios (\ref{ratios1}-\ref{ratios3}) are predictions of our model, to
be verified by future data.  More elaborate models, taking into
account non-thermal $J/\psi$ production, are described in
Refs. \cite{jpsi1,jpsi2}.

More detailed information about the hadron production is, of course,
contained in the transverse-momentum spectra. Several calculations
have been performed recently in the hydrodynamic approaches
\cite{huo,tea,jap,lcso}. In the framework of the thermal models such a
calculation is much more involved than the analysis of the abundances,
since it requires the implementation \cite{wbwf,wfwbmm} of the cascade
processes according to the formulas of the Appendix.  The results are
shown in Fig. \ref{f1}. The upper part displays the spectra of pions,
kaons, antiprotons (already presented in \cite{wbwf}
\footnote{In Ref. \cite{wbwf} the preliminary data for $\overline{p}$
\cite{harris} were presented that exclude the feeding from the weak
decays through the use of the HIJING model. The correction for weak
decays results in about 20\% reduction of the normalization of the
spectrum. In the present paper the official data are presented
\cite{starantip}, which include the full feeding from weak
decays.}), the $\phi$ mesons, and the $K^*(892)^0$ mesons.  With the
parameters fixed in Ref. \cite{wbwf} with help of the spectra of
pions, kaons, and antiprotons, we now obtain the $p_\perp$-spectrum of
the $\phi$ mesons, which agrees very well with the recently-reported
measurement \cite{yama}. The model curve crosses five out of the nine
data points; one should also bear in mind that the systematic
experimental errors are expected at the level of about 20\%
\cite{yama}.  The experimental ratios for the most-central events
\cite{yama} agree
well with the output of our model:   
\begin{eqnarray}
{\phi \over h^-} &=& 0.019 \hspace{0.4cm} (0.021 \pm 0.001)_{\rm expt.}, 
\nonumber \\
{\phi \over K^-} &=& 0.15  \hspace{0.4cm} (0.10-0.16)_{\rm expt.}.
\label{phi}
\end{eqnarray}
\begin{figure}[t]
\epsfysize=13.4cm
\centerline{\mbox{\epsfbox{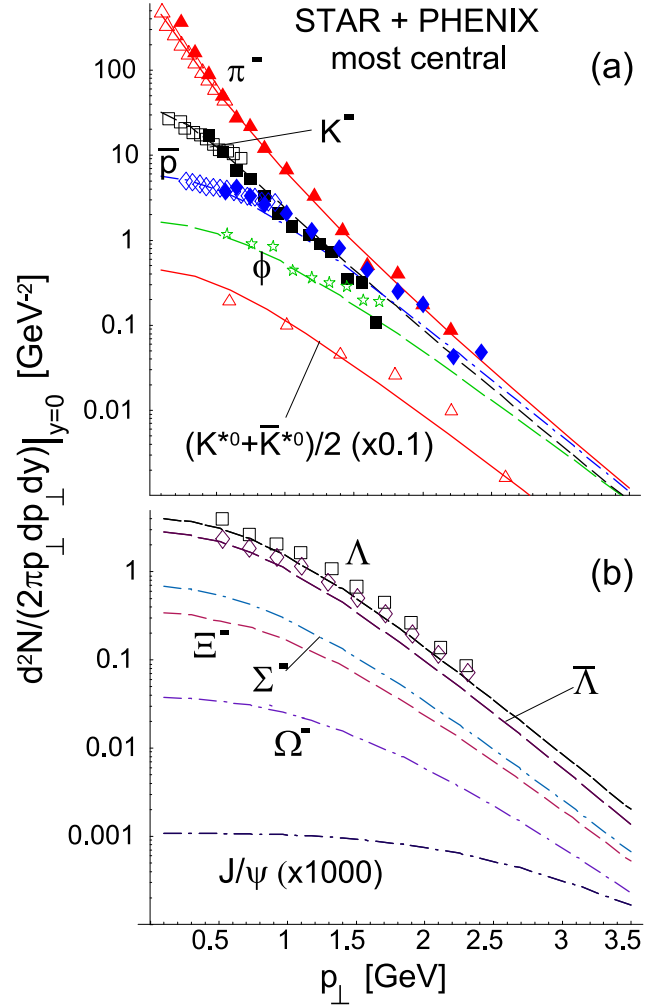}}}
\caption{ (a) The $p_{\perp}$-spectra at midrapidity of pions, kaons,
antiprotons, the $\phi$-mesons, and the $K^*(892)^0$ mesons, and (b)
of $\Lambda$, ${\bar \Lambda}$, $\Sigma^-$, $\Xi^-$ , $\Omega^-$, and
$J/\psi$.  The open  and filled symbols describe the STAR and PHENIX
highest-centrality data, respectively, for $Au+Au$ at $s^{1/2}= 130$
GeV A.  All theoretical curves and the data are absolutely normalized.
The two expansion parameters of the model were fitted to the spectra
of pions, kaons, and antiprotons in Ref. [1]. The model curves include
full feeding from the weak decays. The data come from
Refs. [2,3,31-33]. The STAR data for $\pi^-,K^-,\phi$ and $K^*$'s are
preliminary. The data include full feeding from weak decays.}
\label{f1}
\end{figure}
We recall that the $\phi$ meson deserves a particular attention in
relativistic heavy-ion collisions. It serves as a very good
``thermometer'' of the system, since its interaction with the hadronic
environment is negligible. Also, it does not obtain any contribution
from the resonance decays, thus its spectrum reflects directly the
distribution at freeze-out and the flow.

The upper part of Fig. \ref{f1} shows also the averaged spectrum of
$K^*$'s. In this case the model calculation is compared to the very
recently released data \cite{starKstar}.  Once again we observe a very
good agreement between the model curve and the experimental
points. The measured ratios involving the yield of ${K^*}(892)^0$
\cite{starKstar} also agree well with the output of the thermal model:
\begin{eqnarray}
{{\overline{K^{* 0}}} \over  K^{* 0}} &=& 0.90 \hspace{0.4cm} 
(0.92 \pm 0.14)_{\rm expt.}\, , \nonumber \\
{K^{* 0} \over h^{-}} &=& 0.046  \hspace{0.4cm} 
(0.042 \pm 0.004 \pm 0.01)_{\rm expt.} \, , \nonumber \\
{K^{* 0} \over K} &=& 0.33 \hspace{0.4cm}  
(0.26 \pm 0.03 \pm 0.07)_{\rm expt.}\, , \nonumber \\
{\phi \over K^{* 0}} &=& 0.42 \hspace{0.4cm}  
(0.49 \pm 0.05 \pm 0.12)_{\rm expt.}\, .
\end{eqnarray}
As already mentioned in the Introduction, the successful description
of both the yield and the spectrum of $K^*(892)^0$ mesons supports the
concept of the thermal description of hadron production at RHIC, and
brings evidence for small expansion time between chemical and thermal
freeze-outs.  If the $K^*(892)^0$ mesons decayed between chemical and
thermal freeze-out, the emitted pions and kaons would rescatter and
the $K^*(892)^0$ states could not be reconstructed. In addition, if
only a fraction of the $K^*(892)^0$ yield was reconstructed, it would
not agree with the outcome of the thermal analysis which provides the
particle yields at the chemical freeze-out. Thus, the expansion time
between chemical and thermal freeze-out should be smaller than the
$K^*(892)^0$ lifetime, $\tau =$ 4 fm/$c$ \cite{starKstar}.

The lower part of Fig. \ref{f1} shows the spectra of strange baryons,
and the $J/\psi$ meson. The data for $\Lambda$ and
$\overline{\Lambda}$ were taken from Ref. \cite{starLambda}.  The
$p_\perp$-dependence of the spectrum of $\Lambda$ and
$\overline{\Lambda}$ is quite satisfactorily reproduced,
within 20\%,  by the model
calculation.  The other curves in the figure are our predictions.

\begin{figure}[t]
\epsfysize=9.5cm
\centerline{\mbox{\epsfbox{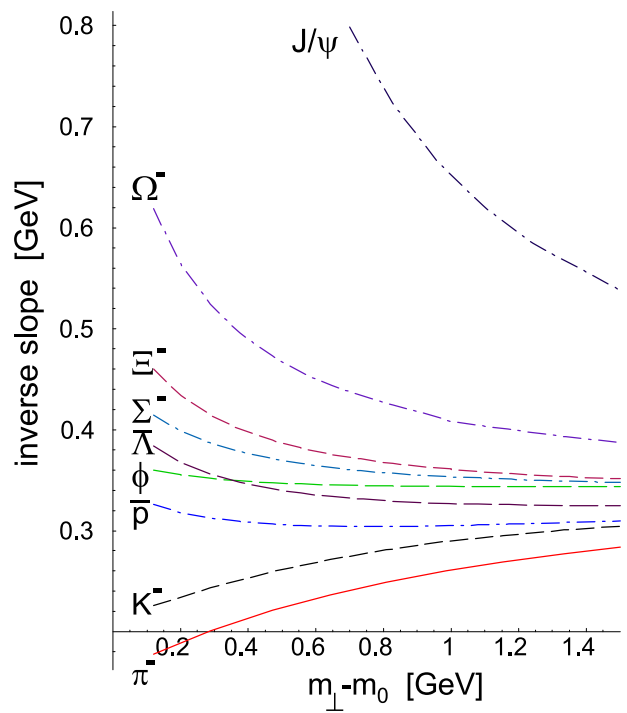}}}
\caption{The inverse-slope parameters for various particles,
calculated as a function of the transverse mass, $m_\perp$. The
constant $m_0$ is the mass of the given particle.}
\label{f2}
\end{figure}

In Fig. \ref{f2} we give the inverse slope parameters for various
particles, defined as
\begin{equation}
\lambda_i=-\left [ \frac{d}{dm_\perp}
\ln  \left( \left.   \frac{dN_i}{2 \pi m_\perp dm_\perp dy}\right|_{y=0}\right)
\right]^{-1}.
\end{equation}
This definition reduces to a constant for the exponential spectrum,
$\exp (-m_\perp/\lambda)$.  As can be seen from the figure, the
inverse slopes depend strongly on the value of $m_\perp$ in the region
of interest.  The same conclusion has been reached in Ref. \cite{SSH}.
Thus, at least from the viewpoint of our model, it is impossible to
describe the spectra in terms of a single slope parameter, as is most
commonly practiced in the interpretation of the data. Secondly, the
inverse slope increases with the growing mass of the hadron, however
this increase is not linear and it is not possible to parameterize it
with a simple formula. Recall that the spectra are fed by the decays
of resonances, which is an important effect, significantly reducing
the inverse slope at lower values of $m_\perp$ \cite{wfwbmm}. The
$m_\perp$ spectra of pions and kaons are {\em convex}~\footnote{
We use the convention that a function with positive second derivative
is convex.}, whereas for the
remaining particles they are concave, the more the heavier the
particle is.  All curves in Fig. \ref{f2} asymptote to the value
\begin{equation}
\lambda_\infty \simeq
\frac{T}{\sqrt{1+\rho_{\rm max}^2/\tau^2}-\rho_{\rm max}/
\tau}=363{\rm MeV}.
\end{equation}
This formula, specific for our expansion model, is simple to obtain
from Eq. (\ref{Ni}) with $f_i$ given by the exponential form, which is
true at large $m_\perp$, where the feeding from the resonance decays
is absent. As can be seen from Fig. \ref{f2}, the asymptotics is reached
slowly for most of the particles.

\section{Excluded-volume effects}

Our fitted values for the geometric parameters $\tau$ and $\rho_{\rm
max}$ are low, of the order of the size of the colliding nuclei. This
leads to two problems: 1) the values of the HBT radii \cite{hbt} are
too small compared to experiment, and 2) there is little time left for
the system to develop large transverse flow. The problems can be
helped by the inclusion of the excluded-volume (van der Waals)
corrections.  Such effects were realized to be important in the
earlier studies of the particle multiplicities in relativistic
heavy-ion collisions \cite{braunmu,gore0,gore}, where they led to a
significant dilution of system.  In the case of the classical
(Boltzmann) statistics, which is a very good approximation for our
system \cite{mm}, the excluded volume corrections bring in a factor
\cite{gore0}
\begin{equation}
\frac{e^{-P v_i /T}}{1+\sum_j v_j e^{-P v_j /T} n_j},
\label{vdw}
\end{equation}
into the phase-space integrals, 
where $P$ is the pressure, $v_i=4 \frac{4}{3}\pi r_i^3$ 
is the excluded volume for the 
particle of species $i$, and $n_i$ is the density of particles of species $i$.
The pressure is
calculated self-consistently from the equation 
\begin{equation}
P=\sum_i P^0_i(T, \mu_i-P v_i/T)=\sum_i P^0_i(T, \mu_i)e^{-P v_i/T},
\label{pres}
\end{equation}
where $P^0_i$ is the partial pressure of the ideal gas of hadrons 
of species $i$.
For the simplest case where the excluded volumes for all particles are equal, 
$r_i=r$, the correction manifests itself as a common scale factor,
which we denote by $S^{-3}$. The
Frye-Cooper formula can then be written in the form \cite{wbwf}
\begin{eqnarray}
\frac{dN_{i}}{d^{2}p_{\perp }dy} &=&\ \tau ^{3}\int_{-\infty }^{+\infty
}d\alpha _{\parallel }\int_{0}^{\rho _{\max }/\tau }{\rm sinh}  \alpha _{\perp
}d\left( {\rm sinh}  \alpha _{\perp }\right)  \nonumber \\
&&\times \int_{0}^{2\pi }
d\xi \, p\cdot u \, S^{-3} f_{i}\left( p\cdot u\right) ,
\label{dNi}
\end{eqnarray}
where
$p\cdot u=m_{\perp }{\rm cosh} \alpha _{\parallel } {\rm cosh}  \alpha
_{\perp }-p_{\perp }\cos \xi \, {\rm sinh}  \alpha _{\perp }$.
The emergence of the factor $S^{-3}$ in Eq. (\ref{dNi})
may be compensated by rescaling $\rho$ and $\tau$
by the factor $S$. That way the system becomes more dilute and larger in such
a way, that the particle multiplicities and the spectra are left intact.

For our values of the thermodynamic parameters, with $\sum_i P^0_i(T,
\mu_i)=80$MeV/fm$^{3}$, we find $S=1.3$ with $r=0.6$fm, and $S=1.6$
with $r=0.8$fm. Such values of the excluded volumes have been
typically used in other calculations. That way, the increase of the
size parameters at freeze-out of the order of 30-60\% is generated.
Therefore, the inclusion of the excluded-volume corrections opens the
possibility that the problems 1) and 2) may be alleviated.
More detailed analysis will be presented elsewhere.

\section{Conclusions}

The presented model results for the strange hadron production at RHIC
support the idea that also these particles are produced thermally.  No
extra parameters ({\em e.g.}  strangeness suppression factors) are
necessary.  This is an important information concerning the particle
production mechanism at RHIC.  Moreover, the hypothesis of a single
freeze-out of Ref. \cite{wbwf} has been additionally verified with the
available spectra of $\phi$, $\Lambda$,  
$\overline{\Lambda}$, and $K^*$'s. In short, the
thermal model supplied with a suitable parameterization of the
freeze-out hypersurface and velocity ({\em i.e.} the inclusion of the
longitudinal and transverse flow) allows for an efficient and uniform
description of the RHIC spectra.  A further verification of the model
will be provided by the calculation of the HBT correlation radii, as
well as the analysis of non-central events, which are currently being
examined.

We are grateful to Marek Ga\'zdzicki for helpful discussions.

\appendix

\section{}

Let us consider a sequence of the resonance decays. The initial
resonance decouples on the freeze-out hypersurface at the space-time
point $x_N$, and decays after an average time inversely 
proportional to the width, $\tau_N \approx
1/\Gamma_N$.  We follow one of the decay products, formed at the point
$x_{N-1}$. It decays again after a time $\tau_{N-1}$, and so on. 
At the end of
the cascade a particle with the label 1 is formed, which is directly
observed in the experiment. The Lorentz-invariant phase-space density
of the measured particles is (we generalize below the formula from
Ref. \cite{ornik} where a single resonance is taken into account)
\begin{eqnarray}
&&n_{1 }\left( x_{1},p_{1}\right)  =  \nonumber \\
&& \nonumber \\
&& \int \frac{d^{3}p_{2}}{E_{p_{2}}}
B\left( p_{2},p_{1}\right) \int d\tau _{2}\Gamma _{2}e^{-\Gamma _{2}\tau
_{2}} \nonumber \\
&& \hspace{1cm} \times \int d^{4}x_{2}\delta ^{\left( 4\right) }
\left( x_{2}+\frac{p_{2}\tau_{2}}{m_{2}}-x_{1}\right) ...   \nonumber \\
&& \nonumber \\
&&\times \int \frac{d^{3}p_{N}}{E_{p_{N}}}B\left( p_{N},p_{N-1}\right) \int
d\tau _{N}\Gamma _{N}e^{-\Gamma _{N}\tau _{N}} \nonumber \\
&& \hspace{1cm} \times \int d\Sigma _{\mu }\left(
x_{N}\right) \,p_{N}^{\mu }\,\,\delta ^{\left( 4\right) }\left( x_{N}+\frac{%
p_{N}\,\tau _{N}}{m_{N}}-x_{N-1}\right) \nonumber \\ 
&& \hspace{1.5cm} \times \,
f_{N}\left[ p_{N}\cdot u\left(x_{N}\right) \right], 
\nonumber \\
\label{npix1p1}
\end{eqnarray}
where $B(q,k)$ is the probability for a resonance with momentum
$k$ to produce a particle with momentum $q$, namely
\begin{equation}
B(q,k) = {b \over 4 \pi p^*} \delta \left(
{k \cdot q \over m_R}- E^* \right).
\label{B}
\end{equation} 
Here $b$ is the branching ratio
for the particular decay channel, and $p^*$ ($E^*$) is the momentum
(energy) of the emitted particle in the resonance's rest frame.
Integration over all space-time positions gives the momentum
distribution 
\begin{eqnarray}
&& E_{p_1} {dN_1 \over d^3 p_1} = \int d^{4}x_{1}\,n_{1}
\left(x_{1},p_{1}\right)   \nonumber \\
&=&\int \frac{d^{3}p_{2}}{E_{p_{2}}}B\left( p_{2},p_{1}\right) ... 
\int \frac{d^{3}p_{N}}{E_{p_{N}}}B\left( p_{N},p_{N-1}\right) 
\nonumber \\
&&
\times \int d\Sigma_{\mu }\left( x_{N}\right) \,p_{N}^{\mu }\,\,f_{N}
\left[ p_{N}\cdot u\left(x_{N}\right) \right], \nonumber \\
\label{npip1}
\end{eqnarray}
which should be used in a general case.
Note that the dependence on the widths $\Gamma_k$ has disappeared,
reflecting the fact that it is not relevant when or where 
the resonances decay.

A simplification follows 
if the element of the freeze-out hypersurface is proportional to the
four-velocity (this is exactly the case considered in our model
defined by Eqs. (\ref{tau}) and (\ref{umu})),
\begin{equation}
d\Sigma_{\mu }(x_{N})=d\Sigma(x_{N}) \,u_\mu(x_{N}).
\label{A4}
\end{equation}
Then
\begin{eqnarray}
&& E_{p_1} {dN_1 \over d^3 p_1}=\int d\Sigma \left(
x_{N}\right) \int \frac{d^{3}p_{2}}{E_{p_{2}}}B\left( p_{2},p_{1}\right)
... \nonumber \\
&&\times
\int \frac{d^{3}p_{N}}{E_{p_{N}}}B\left( p_{N},p_{N-1}\right) p_{N}
\,\cdot u\left( x_{N}\right) \,f_{N}\left[ p_{N}\cdot u\left( x_{N}\right)
\right]  \nonumber \\
&=&\int d\Sigma \left( x_{N}\right)
p_{1}\,\cdot u\left( x_{N}\right)
\,f_{1}\left[ p_{1}\cdot u\left( x_{N}\right) \right],
\label{A5}
\end{eqnarray}
where we have introduced the transformation
\begin{eqnarray}
& & p_{k-1}\,\cdot u\left( x_{N}\right) \,f_{k-1}\left[ p_{k-1}\cdot
u\left( x_{N}\right) \right] \nonumber \\
& & =\int \frac{d^{3}p_{k}}{E_{p_{k}}}B\left(
p_{k},p_{k-1}\right) p_{k}\,\cdot u\left( x_{N}\right) \,f_{k}\left[
p_{k}\cdot u\left( x_{N}\right) \right], \nonumber \\
\label{trsim}
\end{eqnarray}
which can be used (step by step along the cascade) to calculate the
distribution of the measured particles. In the fluid local-rest-frame,
most convenient in the numerical calculation, we have
$u^\mu(x_N)=(1,0,0,0)$, and the transformation (\ref{trsim}) reduces
to the form discussed in \cite{wfwbmm}.  A technical simplification
relies in the fact that in Eq. (\ref{A5}) the space-time integration over
the hypersurface is performed at the end, consequently the momentum
integration in Eq. (\ref{trsim}) preserves the full spherical
symmetry.  That way, the momentum integrals are one-dimensional and
the numerical procedure is very fast. On the other hand, in the
general case of Eq. (\ref{npip1}) the integration over the
hyper-surface has to be done first, reducing the symmetry of the
following momenta integrals.  The symmetry is cylindrical, thus the
momentum integrals are two-dimensional.  Moreover the integrands have
an integrable singularity, which make the numerical procedure more
involved \cite{ornik,sollfrank}.


\begin{thebibliography}{99}

\bibitem[\ast]{grant} Supported in part by the Polish State Committee for
Scientific Research, grant 2 P03B 09419.

\bibitem{wbwf} W. Broniowski and W. Florkowski,
Phys. Rev. Lett. {\bf 87}, 272302 (2001).

\bibitem{yama} E. T. Yamamoto, hep-ph/0112017.

\bibitem{starLambda} C. Adler et al., STAR Collaboration, nucl-ex/0203016.

\bibitem{starKstar} P. Fachini, STAR Collaboration, nucl-ex/0203019.

\bibitem{rafmul} J. Rafelski and B. M\"uller, Phys. Rev. Lett. {\bf
48}, 1066 (1982).

\bibitem{kochmulraf} P. Koch, B. M\"uller, and J. Rafelski,
Phys. Rep. {\bf 142}, 167 (1986).

\bibitem{raf} J. Rafelski and J. Letessier, hep-ph/0112027.

\bibitem{singh} C. P. Singh, Phys. Rep. {\bf 236}, 147 (1993).

\bibitem{pbmsps} P. Braun-Munzinger, I. Heppe, and J.~Stachel,
Phys. Lett. B {\bf 465 }, 15 (1999).

\bibitem{pbmrhic} P. Braun-Munzinger, D. Magestro, K. Redlich, and J. Stachel,
Phys. Lett. B {\bf 518}, 41 (2001).

\bibitem{wfwbmm} W. Florkowski, W. Broniowski, and M. Michalec,
Acta Phys. Pol. B {\bf 33}, 761 (2002).

\bibitem{heinzr} U. Heinz, Nucl. Phys. A {\bf 661}, 140c (1999).

\bibitem{rafelski} J. Rafelski and J. Letessier,
Phys. Rev. Lett. {\bf 85}, 4695 (2000).

\bibitem{PDG}  Particle Data Group, Eur. Phys. J. C {\bf 15}, 1  (2000).

\bibitem{hagedorn} R. Hagedorn, CERN preprint No. CERN-TH.7190/94 (1994),
and references therein.

\bibitem{bjorken} J. D. Bjorken, Phys. Rev. D {\bf 27}, 140 (1983).

\bibitem{baym} G. Baym, B. Friman, J.-P. Blaizot, M. Soyeur, and W.~Czy\.z,
Nucl. Phys. A {\bf 407}, 541 (1983). 

\bibitem{Kolya} P. Milyutin and N. N. Nikolaev, Heavy Ion Phys {\bf 8}, 
333 (1998); V. Fortov, P. Milyutin, and N. N. Nikolaev,
JETP Lett. {\bf 68}, 191 (1998).

\bibitem{siemens} P. J. Siemens and J. Rasmussen, Phys. Rev. Lett. {\bf 42},
880 (1979); P. J. Siemens and J. I. Kapusta, Phys. Rev. Lett. {\bf 43},
1486 (1979).

\bibitem{SSH} E. Schnedermann, J. Sollfrank, and U. Heinz,
Phys. Rev. C {\bf 48}, 2462 (1993).

\bibitem{BL} T. Cs\"{o}rg\H{o} and B. L\"{o}rstad, Phys. Rev. C {\bf
54}, 1390 (1996).

\bibitem{Rischke} D. H. Rischke and M. Gyulassy, Nucl. Phys. A {\bf
697}, 701 (1996); Nucl. Phys. A {\bf 608}, 479 (1996).

\bibitem{SH} R. Scheibl and U. Heinz, Phys. Rev. C {\bf 59}, 1585
(1999).

\bibitem{bugaev} K. A. Bugaev, Nucl. Phys. A {\bf 606}, 559 (1996).

\bibitem{csernai} L. P. Csernai, Zs. I. L\'az\'ar, and D. Moln\'ar,
Heavy Ion Phys. {\bf 5}, 467 (1997).

\bibitem{neymann} J. J. Neymann, B. Lavrenchuk, and G. Fai,
Heavy Ion Phys. {\bf 5}, 27 (1997).

\bibitem{magas} V. K. Magas {\em et al.}, Nucl. Phys. A {\bf 661}, 596c (1999).


\bibitem{CF2} F. Cooper, G. Frye, and E. Schonberg,
Phys. Rev. D {\bf 11}, 192 (1975).

\bibitem{biro} T. S. Bir\'o, Phys. Lett. B {\bf 474}, 21 (2000);
Phys. Lett. B {\bf 487}, 133 (2000).

\bibitem{lhs} K.~S.~Lee, U. Heinz, and E. Schnedermann, Zeit. f. Phys.
C {\bf 48}, 525 (1990). 

\bibitem{velko} J. Velkovska, PHENIX Collaboration, 
Nucl. Phys. A {\bf 698}, 507 (2002).

\bibitem{harris} J. Harris, STAR Collaboration, talk presented at QM2001.

\bibitem{starantip} C. Adler et al., STAR Collaboration, Phys. Rev.
Lett. {\bf 87}, 262302 (2001).

\bibitem{CF1} F. Cooper and G. Frye, Phys. Rev. D {\bf 10}, 186 (1974).

\bibitem{hag1} W. Broniowski and W. Florkowski, Phys. Lett. B {\bf 490},
223 (2000).

\bibitem{hag2} W. Broniowski, in Proc. of Few-Quark Problems, Bled, 
Slovenia, July 8-15, 2000, eds. B. Golli, M. Rosina, and S. \v Sirca, 
p. 14, hep-ph/0008122.

\bibitem{gaz1} M. Ga\'zdzicki and M. I.  Gorenstein, Phys. Rev. Lett.
{\bf 83}, 4009 (1999).

\bibitem{gaz2} K. A. Bugaev, M. Ga\'zdzicki, and M. I.  Gorenstein, 
Phys. Lett. B {\bf 523}, 255 (2001).

\bibitem{jpsi1} L. Grandchamp and R. Rapp, Phys. Lett. B {\bf 523}, 60
(2001).

\bibitem{jpsi2} P. Braun-Munzinger and J. Stachel, Nucl. Phys. A
{\bf 690}, 119 (2001); Phys. Lett. B {\bf 490}, 196 (2000).

\bibitem{huo} P. Huovinen, P. F. Kolb, U. Heinz, P. V. Ruuskanen, and 
S. A. Voloshin, Phys. Lett. B {\bf 503}, 58 (2001).

\bibitem{tea} D. Teaney, J. Lauret, and E. V. Shuryak,
Phys. Rev. Lett. {\bf 86}, 4783 (2001);  
Nucl. Phys. A {\bf 698}, 479 (2002).

\bibitem{jap} T. Hirano, Phys. Rev. C {\bf 65}, 011901 (2002);
T. Hirano, K. Morita, S. Muroya, and C. Nonaka, nucl-th/0110009.

\bibitem{lcso} A. Ster and T. Cs\"org\H{o}, hep-ph/0112064.

\bibitem{hbt} C. Adler et al., STAR Collaboration, Phys. Rev. Lett.
{\bf 87}, 082301 (2001).

\bibitem{braunmu}  P. Braun-Munzinger, J. Stachel, J. P. Wessels, and N. Xu,
Phys. Lett. B {\bf 344}, 43 (1995); Phys. Lett. B {\bf 365}, 1 (1996).

\bibitem{gore0}  G. D. Yen,  M. I. Gorenstein, W. Greiner, and S. N. Yang,
Phys. Rev. C {\bf 56}, 2210 (1997).


\bibitem{gore}  G. D. Yen and M. I. Gorenstein, Phys. Rev. C {\bf 59}, 2788
(1999).

\bibitem{mm} M. Michalec, PhD Thesis, nucl-th/0112044.

\bibitem{ornik} J. Bolz, U. Ornik, M. Pl\"umer, B.R. Schlei, and
R.M.~Weiner, Phys. Rev. D {\bf 47}, 3860 (1993).

\bibitem{sollfrank} 
J. Sollfrank, P. Koch, and U. Heinz, Phys. Lett. B {\bf 252}, 256 (1990).




\end{thebibliography}
\end{document}